\input amstex
\documentstyle{amsppt}
{\catcode`\@=11\gdef\logo@{}}
\TagsOnRight
\loadbold
\pageheight{25 true cm}
\pagewidth{17 true cm}
\document
\centerline{\bf On 3+1 decompositions with respect to an observer field
                                                via differential forms}
\vskip 2cm
\centerline{Mari\'an Fecko ${}^{a)}$}
\centerline{Department of Theoretical Physics, Comenius University}
\centerline{Mlynsk\'a dolina F2, 842 15 Bratislava, Slovakia}
\vskip 2cm
{\bf Abstract}
\vskip 1cm
3+1 decompositions of differential forms on a Lorentzian manifold
($M,g;+ - - -$) with respect to arbitrary observer field and the decomposition
of the standard operations acting on them are studied, making use of the
ideas of the theory of connections on principal bundles. Simple
explicit general formulas are given as well as their application to the Maxwell
equations.
\vskip 2cm
PCAS : 02.40.-k, \ \   04.20.-q, \ \ 03.50.De
%
%
\vskip 2cm
\leftline{running title : On 3+1 decompositions via differential forms}
\vskip 2cm
{\bf I. Introduction}
\vskip 0.5cm
\indent
There is rather extensive literature devoted to 3+1 split of the laws
of physics
in curved spacetime (cf. [1] for a review and also the references therein).
According to Sec. 2. of [1] there are two rather different methods
available to cope with the problem, viz. the congruence method and the
hypersurface one.
\newline \indent
Here we present a systematic method of 3+1 split within the {\it congruence}
method using the language of {\it differential forms} on both (4 and 3+1)
levels.
\newline \indent
The use of forms within the 3+1 decomposition program can be traced back to
the classical paper on geometrodynamics [2] (p.581), where it was applied,
however,
in the framework of the {\it hypersurface} method (cf. also [3], pp. 93-94);
the time serves there as a parameter labeling the spacelike hypersurfaces and
the time derivative of a form is interpreted as a differentiation with respect
to a parameter.
\newline \indent
The observer field approach similar to ours can be found in [4] (pp.
193-197). What we add here is the introduction of the (simply realized)
operator hor (cf. Sec.3) and spatial exterior derivative within the general
congruence approach (Sec. 4b). These objects turn out to be very convenient
to manipulate with and enable one to derive very simple and at the same time
general decomposition formulas and rules.
\newline \indent
The following data are assumed in the article :
a 4-dimensional Lorentzian ($g$ of the signature + - - -) manifold $M$ with
orientation ( $\equiv$ spacetime), and an {\it observer} (velocity)
{\it field}, i.e. a future oriented vector field on $M$ obeying
$$g(V,V) \equiv \mid \mid V {\mid \mid}^2 =1 \tag 1.1$$
(the integral curves of $V$ provide then the congruence of proper-time
parametrized worldlines of observers).
\newline \indent
All the constructions are in fact local, i.e. it is enough that the objects
mentioned above are available only in a domain $\Cal U\subset M$ rather than
globally and consequently no global properties of $M$ are assumed.
\vskip 1cm
\noindent
{\bf II. The decomposition of forms}
\vskip 0.5cm
\indent
For any vector field $W$ let us define the following standard operations on
forms :
\newline
$$i_W : {\Omega}^p(M) \rightarrow  {\Omega}^{p-1}(M)$$
$$j_W : {\Omega}^p(M) \rightarrow  {\Omega}^{p+1}(M)$$

$$i_W\alpha (U,...) := \alpha (W,U,...) \tag 2.1$$
$$j_W\alpha := \tilde W \wedge \alpha \hskip 1cm \tilde W \equiv g(W,.)
\equiv {\flat}_gW \tag 2.2$$
Then the following identity holds :
$$j_Wi_U + i_Uj_W = g(U,W) , \tag 2.3$$
in particular
$$j_Vi_V + i_Vj_V =  \ 1 \equiv \text{identity on} \ \Omega (M).
                                                                \tag 2.4$$
Further, introducing
$$\Cal P := i_Vj_V \hskip 1cm \Cal Q := j_Vi_V \tag 2.5$$
one checks easily that they represent the set of {\it projection operators}
on  ${\Omega}^p(M)$, i.e.
$${\Cal P}^2 = \Cal P \hskip 0.5cm {\Cal Q}^2 = \Cal Q \hskip 0.5cm
     \Cal P\Cal Q = 0 = \Cal Q \Cal P \hskip 0.5cm \Cal P + \Cal Q = 1
                                                                 \tag 2.6$$
Then for any $\alpha \in {\Omega}^p(M)$ one has
$$\alpha = (\Cal Q + \Cal P)\alpha = \tilde V \wedge i_V\alpha +
                                                    i_Vj_V\alpha \tag 2.7$$
i.e. one obtains the {\it decomposition}
$$\alpha = \tilde V \wedge \hat s + \hat r \tag 2.8$$
where
$$\hat s \equiv i_V\alpha \hskip 1cm \hat r \equiv i_Vj_V \alpha \tag 2.9$$
\vskip 1cm
\noindent
{\bf III. Operator hor and spatial forms}
\vskip 0.5cm
At any point $m \in M$ we define
{\it vertical} \ (instantaneous time) direction - parallel to $V$ and
{\it horizontal} (instantaneous 3-spa\-ce) directions - perpendicular to $V$.
Then for any vector there is the unique decomposition
$$U=U_{\mid \mid} + U_{\perp} \equiv \text{ver} \ U +
                     \text{hor} \ U$$
and one can define (in the spirit of the theory of connections on principal
bundles, cf. [5] )
$$(\text{hor} \ \alpha ) (U,W,...) := \alpha (\text{hor} \ U,\text{hor} \
  W,...) \tag 3.1$$
It turns out (cf. Appendix A) that this operation is realized explicitly as
$$\text{hor} \ \alpha = i_Vj_V\alpha \equiv \Cal P \alpha \equiv \hat r
                                                                \tag 3.2$$
so that the decomposition in \thetag{2.8} can be rewritten also as
$$\alpha = \tilde V \wedge i_V\alpha + \text{hor} \ \alpha \tag 3.3$$
We also introduce the space of purely {\it spatial} (horizontal) {\it p-forms}
by
$$\hat {\Omega}^p(M) := \{ \alpha \in {\Omega}^p(M) \ | \ \alpha =
                \text{hor} \ \alpha \} \tag 3.4$$
(i.e. $\hat s = 0$ in the decomposition \thetag {2.8})
and the Cartan algebra of spatial forms
$$\hat \Omega (M) := {\oplus}_p \hat {\Omega}^p(M) \tag 3.5$$
(it is closed with respect to $\wedge$).
One readily verifies (inserting arguments and using the definition \thetag
{3.1}) that the projection operator $\Cal P \equiv$ hor is compatible with
the algebra structure in $\Omega (M)$
$$\text{hor} (\alpha + \lambda \beta) = \text{hor} \ \alpha + \lambda \
             \text{hor} \ \beta
             \hskip 2cm
             \alpha, \ \beta \ \in \Omega (M), \lambda \in \Bbb R \tag 3.6$$
$$\text{hor} \ (\alpha \wedge \beta) = \text{hor} \ \alpha \wedge
             \text{hor} \ \beta
             \hskip 2cm
             \alpha, \beta \ \in \Omega (M) \tag 3.7$$
i.e.
$$\text{hor} \ : \ \Omega (M) \rightarrow \text{Im} \ \text{hor} \equiv
                   \hat \Omega (M)  \le \Omega (M) \tag 3.8$$
is an {\it endomorphism} of the Cartan algebra $\Omega (M)$.
From \thetag{3.3} we obtain useful criterion
$$\alpha = \text{spatial form} \hskip 1cm \Leftrightarrow \hskip 1cm
                                                     i_V \alpha =0 \tag 3.9$$
Then we see that $\hat r , \hat s$ in the decomposition \thetag{2.8} are
spatial (\thetag {2.9} plus $i_Vi_V = 0$).
\newline
Note : If a local orthonormal frame field
$e_a \equiv (e_0\equiv V,e_i)$ and its dual $e^a \equiv (e^0\equiv
\tilde V,e^i)$, are used and if
$$\alpha = \frac{1}{p!}{\alpha}_{a ...b}e^a\wedge ... \wedge e^b, \tag 3.10$$
then the decomposition \thetag {2.8} is just the split into two parts
which do and do not contain respectively the basis 1-form $e^0\equiv \tilde V,$
i.e.
$$\alpha = e^0\wedge \hat s +\hat r \tag 3.11$$
where $\hat s, \hat r$, being spatial, do not already contain the local ``time"
basis 1-form $e^0$, but rather only the ``spatial" basis 1-forms $e^i$;
explicitly
$$\hat s = \frac{1}{(p-1)!}{\alpha}_{0i ... j} \undersetbrace {p-1} \to
  {e^i\wedge ... \wedge e^j}
  \hskip 1.5cm
  \hat r = \frac{1}{p!}{\alpha}_{k ... j} \undersetbrace p \to
  {e^k \wedge ... \wedge e^j} \tag 3.12$$
\vskip 1cm
\noindent
{\bf IV. The decomposition of the operations on forms}
\vskip 0.5cm
According to \thetag {2.8} any form on $(M,g,V)$ can be 3+1 decomposed as
$$\alpha = \tilde V \wedge \hat s + \hat r $$
so that the full information about $\alpha \in {\Omega}^p(M)$ is encoded
(in an observer dependent way) into a {\it pair} of
{\it spatial} forms
$\hat s \in  \hat {\Omega}^{p-1}(M)$ and $\hat r \in \hat {\Omega}^p(M)$.
In this section we perform the decomposition of the standard {\it operations}
on forms, viz. the {\it Hodge star} $*$ and the {\it exterior derivative}
$d$ (some other important operators are then easily obtained by their
combinations). By this we mean to introduce some ``spatial" operations
(acting directly on $\hat s, \hat r$ and dependent on the observer) producing
the same effect as does the given operator acting on $\alpha$.
\vskip 1cm
\noindent
{\bf IVa. The Hodge star }
\vskip 0.5cm
The horizontal subspace of a tangent space at each point inherits natural
metric tensor $\hat h$ (with signature + + + by definition, i.e.
$g=\tilde V \otimes \tilde V - \hat h$) and
orientation (a spatial frame $(e_1,e_2,e_3)$ is declared to be right-handed if
$(V\equiv e_0,e_1,e_2,e_3)$ is right-handed). These data are just enough for
the unique {\it spatial Hodge} operator
$$\hat * := *_{\hat h} : \hat {\Omega}^p(M) \rightarrow
                      \hat {\Omega}^{3-p}(M) \tag 4a.1$$
(it is to be applied only on spatial forms). Using the operator
$$\hat \eta \alpha := {(-1)}^{\text{deg}\alpha}\alpha$$
one readily computes (cf. Appendix B) that the decomposition of the ``full"
Hodge star reads
$$*(\tilde V \wedge \hat s + \hat r) = \tilde V \wedge \hat * \hat r
+ \hat * \hat \eta \hat s \tag 4a.2$$
As an example, applying this to $1\in {\Omega}^0(M)$ ($\hat s
=0, \hat r=1$) results in the decomposition of the 4-volume form
$$*1 \equiv \omega = \tilde V \wedge \hat * 1 =: \tilde V \wedge \hat \omega
                                                                \tag 4a.3$$
where
$$\hat \omega := \hat * 1 \tag 4a.4$$
is the {\it spatial volume form}. In local orthonormal right-handed coframe
field $e^a$ it is just
$$\omega \equiv e^0 \wedge e^1 \wedge e^2 \wedge e^3 = e^0 \wedge
 (e^1 \wedge e^2 \wedge e^3) \equiv \tilde V \wedge \hat \omega \tag 4a.5$$
\vskip 1cm
\noindent
{\bf IVb. The exterior derivative }
\vskip 0.5cm
Let $\hat b$ be a spatial form, $\Cal D$ a spatial ($\equiv$ horizontal) domain
(i.e. the domain of any possible dimension with the property that any
vector tangent to it is horizontal). Then
$$ \alignat2 \int_{\Cal D} d\hat b \
  &\overset1.\to= \int_{\partial\Cal D}\hat b &&\text{due to Stokes' theorem}\\
  &\overset2.\to= \int_{\Cal D} \text {hor} \ d\hat b \equiv \int_{\Cal D}
        \hat d \hat b \qquad\qquad &&\text{since $\Cal D$ is horizontal}\\
                                                          \endalignat$$
$\Rightarrow$
$$\int_{\Cal D} \hat d \hat b = \int_{\partial\Cal D} \hat b \tag 4b.1$$
where we introduced the {\it spatial exterior derivative}
$$\hat d : \hat {\Omega}^p(M) \rightarrow \hat {\Omega}^{p+1}(M)$$
$$\hat d := \text{hor} \ d \equiv i_Vj_V \ d \tag 4b.2$$
(exactly like the {\it covariant} exterior derivative of forms on principal
bundle with connection). Thus for spatial forms and domains the ``full"
operator $d$ in the Stokes formula can be replaced by $\hat d$. This means
that $\hat d$ and $\hat *$ provide the basic building blocks for the
``3-dimensional vector analysis" operations, being the natural
generalizations of div, curl etc. used in Minkowski space (div $\sim \hat *
\hat d \hat *$, curl $\sim \hat * \hat d$, ...). We emphasize that the
validity of the {\it spatial Stokes formula} \thetag {4b.1} for $\hat d$
is essential for the usefulness and naturality
of the latter, e.g. as a means to relate the usual differential
3+1 laws to the corresponding integral ones (like $\text{div}\bold B =0
\leftrightarrow \oint \bold B .d\bold S =0$).
\newline \indent
So our task now is to express the action of the ``full" $d$ operator in terms
of $\hat d$ (and possibly some other ones) directly on $\hat s, \hat r$
present in the decomposition \thetag {2.8} of $\alpha$ . We have
$$d \alpha = d\tilde V \wedge \hat s - \tilde V \wedge d \hat s + d\hat r$$
so that we are to focus our attention to two particular issues, viz.
$d$ of $\tilde V$ and $d$ of a spatial form.
\newline \noindent
The decomposition of the 2-form $d\tilde V$ according to \thetag{2.8}
results in
$$d\tilde V=\tilde V \wedge \hat a + \hat y \tag 4b.3$$
with $\hat a \in {\hat \Omega}^1(M), \hat y \in {\hat \Omega}^2(M)$. The
forms $\hat a, \hat y$ are the {\it kinematical characteristics} of the
observer field $V$, which can be easily extracted from any given
$V$ using \thetag {2.9}. Their physical meaning is discussed in Appendix C. It turns out
(see also [6], [7], [8]) that $\hat a$ equals to the {\it acceleration 1-form}
$$\hat a = g({\nabla}_VV,.) \equiv g(a,.) \equiv \tilde a \tag 4b.4$$
($a \equiv  {\nabla}_VV$ is the {\it acceleration field} corresponding
to $V$) and the 2-form $\hat y$, the {\it vorticity form}
(tensor) is the measure of the (non)integrability of the spatial
(horizontal) distribution, i.e. it encodes whether or not the instantaneous
3-spaces mesh together to form a (local) spatial 3-domain $\Cal D$
(or, equivalently, whether or not the {\it time synchronization} is
possible). These properties of $\hat a$ and $\hat y$ are reflected in the
terminology : $V$ is said to be {\it geodesic} if $\hat a =0$, {\it
irrotational} or {\it time-synchronizable} if $\hat y=0$ and {\it
proper-time synchronizable} if both $\hat a$ and $\hat y$ vanish (then
$V={\partial}_t,\tilde V =dt$ in adapted coordinates).
\newline \indent
The computation of the action of $d$ on a spatial form, as well as on a
general form $\alpha$ then, is performed in Appendix D and the result reads
$$d(\tilde V \wedge \hat s + \hat r) = \tilde V \wedge (- \hat d \hat s +
  {\Cal L}_V \hat r + \hat a \wedge \hat s) + (\hat d \hat r + \hat y \wedge
                                                          \hat s) \tag 4b.5$$
The formula \thetag{4b.5} gives the desired 3+1 decomposition of the ``full"
$d$ operator. Notice the explicit occurrence of both kinematical
characteristics $\hat a$ and $\hat y$.
\newline \indent
The spatial exterior derivative $\hat d$ shares some important properties
with the "full" $d$. In particular, it {\it is} the {\it graded derivation} of
the spatial Cartan algebra $\hat \Omega (M)$. Indeed, according to \thetag
{3.7} we have
$$\hat d (\hat r\wedge \hat R)= \text{hor} \ (d\hat r \wedge \hat R +
  \hat \eta \hat r \wedge d\hat R ) =
 \hat d \hat r \wedge \undersetbrace \hat R \to {\text{hor} \ \hat R} +
 \hat \eta \undersetbrace \hat r \to {\text{hor} \ \hat r} \wedge \hat d \hat R
  = \hat d \hat r \wedge \hat R + \hat \eta \hat r \wedge \hat d \hat R
  \tag 4b.6$$
On the other hand, it is {\it not nilpotent} in general, but rather
(see Appendices D and G)
$$\hat d \hat d \hat b = - \hat y \wedge {\Cal L}_V \hat b \hskip 2cm
               \hat b \in \hat \Omega (M) \tag 4b.7$$
holds.
This may seem to contradict \thetag {4b.1}, since the ("full") exterior
derivative can be
uniquely defined by the ("full") Stokes formula [9]
(and it {\it is} then nilpotent due to the nilpotence of the boundary operator).
The situation can be clarified as follows : for {\it any} domain $\Cal D$
and $\alpha \in \Omega (M)$ one has
$$\int_{\Cal D} dd\alpha = \int_{\partial \partial \Cal D} \alpha = 0 \tag
                                                                    4b.8$$
(since $\partial \partial =0$) which leads to $dd\alpha =0$ identically,
i.e. $d$ {\it
is} nilpotent. For a {\it spatial} domain $\Cal D$ and {\it spatial} form
$\hat b$ one has similarly
$$\int_{\Cal D} \hat d \hat d \hat b = \int_{\partial \partial \Cal D} \hat
                                                          b = 0 \tag 4b.9$$
This does not mean, however, that $\hat d \hat d \hat b =0$ identically,
now, but rather $\hat d \hat d \hat b $ should vanish upon restriction to
any {\it spatial} $\Cal D$. The only nontrivial cases are for the dimension
of $\Cal D$ being 3 or 2. For dim $\Cal D$ = 3, $\hat y \neq 0$ (and thus
$\hat d \hat d \hat b \neq 0$ due to \thetag {4b.7}) means (via Frobenius
theorem) that spatial $\Cal D$ (to be used in \thetag {4b.9}) does not
exist at all. For dim $\Cal D $ = 2 we have $\hat b$ = function $\equiv \
f$ and the question is, whether $(Vf)\hat y$ vanishes (for any $f$) upon
restriction on any spatial 2-dimensional domain $\Cal D$. This {\it is},
however, the case as a result of $\hat y$ being the measure of
nonintegrability (the bracket of any two vectors tangent to $\Cal D$ is
trivially again tangent to $\Cal D$). Thus there is {\it no conflict}
between \thetag {4b.1} and \thetag {4b.7}.
\vskip 1cm
\noindent
{\bf V. Matrix notation}
\vskip 0.5cm
For the computation of more complex expressions (e.g. the
{\it codifferential} in \thetag {5.3}) it is quite useful to introduce matrix
realization of the operators. If the decomposition \thetag {2.8} of
$\alpha$ is represented by a column
$$\alpha \equiv \tilde V \wedge \hat s + \hat r \ \leftrightarrow \
\left( \matrix \hat s  \\ \hat r \endmatrix \right)$$
then e.g.
$$*(\tilde V \wedge \hat s + \hat r) =
                  \tilde V \wedge \hat * \hat r + \hat * \hat \eta \hat s
      \leftrightarrow
      \left( \matrix \hat * \hat r  \\ \hat * \hat \eta \hat s \endmatrix
      \right) \equiv \
      \left( \matrix 0 & \hat * \\ \hat * \hat \eta & 0 \endmatrix \right)
      \left( \matrix \hat s  \\ \hat r \endmatrix \right)$$
so that
$$* \ \leftrightarrow \
    \left( \matrix 0 & \hat * \\ \hat * \hat \eta & 0 \endmatrix \right). $$
For the exterior derivative we obtain similarly
$$      d(\tilde V \wedge \hat s + \hat r)
    \leftrightarrow
        \left( \matrix - \hat d \hat s + {\Cal L}_V \hat r + \hat a \wedge \hat
             s \\ \hat d \hat r + \hat y \wedge \hat s \endmatrix \right)
    \equiv
        \left( \matrix -\hat d + \hat a & {\Cal L}_V \\ \hat y &
                 \hat d \endmatrix \right)
        \left( \matrix \hat s \\ \hat r \endmatrix \right)$$
or
$$d \ \leftrightarrow \
    \left( \matrix -\hat d + \hat a & {\Cal L}_V \\ \hat y &
    \hat d \endmatrix \right). $$
In the same sense we can then express also other useful operations in terms of
such matrices; for the sake of convenience we collect them here together :
$$* \ \leftrightarrow \
    \left( \matrix 0 & \hat * \\ \hat * \hat \eta & 0 \endmatrix \right)
    \hskip 1.5cm
    *^{-1} \ \leftrightarrow \
    \left( \matrix 0 & - \hat * \hat \eta \\ \hat * & 0 \endmatrix
    \right)
    \tag 5.1$$
$$\hat \eta \ \leftrightarrow \
    \left( \matrix -\hat \eta & 0 \\ 0 & \hat \eta  \endmatrix \right)
    \hskip 1.5cm
    d \ \leftrightarrow \
    \left( \matrix -\hat d + \hat a & {\Cal L}_V \\ \hat y &
    \hat d \endmatrix \right)
    \tag 5.2$$
$$\delta := *^{-1} d * \hat \eta \ \leftrightarrow \
    \left( \matrix 0 & - \hat * \hat \eta \\ \hat * & 0 \endmatrix
    \right)
    \left( \matrix -\hat d + \hat a & {\Cal L}_V \\ \hat y &
    \hat d \endmatrix \right)
    \left( \matrix 0 & \hat * \\ \hat * \hat \eta & 0 \endmatrix \right)
    \left( \matrix -\hat \eta & 0 \\ 0 & \hat \eta  \endmatrix \right) =$$
  $$\left( \matrix \hat \delta & \hat * (\hat y \wedge \hat * ) \\
    -\hat * {\Cal L}_V \hat *  & -\hat \delta +
    \hat * (\hat a \wedge \hat * \hat \eta)
    \endmatrix \right) \tag 5.3$$
where
$$\hat \delta := {\hat *}^{-1} \hat d \hat * \hat \eta \tag 5.4$$
is the {\it spatial codifferential},
$$i_V \ \leftrightarrow \
    \left( \matrix 0 & 0 \\ 1 & 0 \endmatrix \right)
    \hskip 1.5cm
    j_V \ \leftrightarrow \
    \left( \matrix 0 & 1 \\ 0 & 0 \endmatrix \right)
    \tag 5.5$$
$$\text{hor} = i_V j_V  \ \leftrightarrow \
    \left( \matrix 0 & 0 \\ 1 & 0 \endmatrix \right)
    \left( \matrix 0 & 1 \\ 0 & 0 \endmatrix \right) \ =
    \left( \matrix 0 & 0 \\ 0 & 1 \endmatrix \right)
    \tag 5.6$$
$${\Cal L}_V \equiv i_Vd \ + \ d \ i_V \leftrightarrow \
  \left( \matrix {\Cal L}_V & 0 \\ \hat a & {\Cal L}_V \endmatrix \right)
    \tag 5.7$$
\vskip 1cm
\noindent
{\bf VI. The Maxwell equations}
\vskip 0.5cm
According to the standard conventions on the relationship between the
components of the electromagnetic field 2-form $F \equiv \frac12
F_{ab}e^a\wedge e^b$ ($e_a$ is $g$-orthonormal frame) and the 3-space vectors
of the electric and magnetic fields respectively
$$F_{0\alpha} =E_{\alpha}=E^{\alpha} \hskip 1cm
  F_{\alpha \beta} = -{\epsilon}_{\alpha \beta \gamma}B^{\gamma}
  \equiv -{\epsilon}_{\alpha \beta \gamma}B_{\gamma} \tag 6.1$$
($e_{\alpha}$ is $\hat h$-orthonormal frame; $\alpha, \beta$... run from 1 to 3,
being raised and lowered by the {\it spatial} metric tensor
${\hat h}_{\alpha \beta}$ $\equiv + {\delta}_{\alpha \beta} \equiv -
                                                  {\eta}_{\alpha \beta}$),
one can associate with the electric and magnetic fields the {\it spatial forms}
$$\hat E = E_{\alpha} e^{\alpha} =: {\bold E}.d{\bold r}
  \hskip 1cm
  \hat B = B^{\alpha} dS_{\alpha} =: {\bold B}.d{\bold S}
  \hskip 1cm
  dS_{\alpha} := \frac12 {\epsilon}_{\alpha \beta \gamma}
                             e^{\beta}\wedge e^{\gamma} \tag 6.2$$
Then
$$F= \tilde V \wedge \hat E - \hat B
     \ \leftrightarrow \
     \left( \matrix \hat E \\ -\hat B \endmatrix \right)
     \tag 6.3$$
(so that $\hat s = \hat E, \ \hat r = -\hat B$ here). Similarly the
electric {\it 4-current 1-form} decomposes to
$$j= j_ae^a = j_0e^0 + j_ie^i \equiv \rho \tilde V - \hat j
     \ \leftrightarrow \
     \left( \matrix \rho \\ -\hat j \endmatrix \right)
     \hskip 1cm
     \hat j := j_{\alpha}e^{\alpha} =j^{\alpha}e^{\alpha} \tag 6.4$$
Then
$$*F=\tilde V \wedge (-\hat * \hat B) - \hat * \hat E
     \ \leftrightarrow \
     \left( \matrix -\hat * \hat B \\ -\hat * \hat E \endmatrix \right)
     \tag 6.5$$
$$*j=\tilde V \wedge (-\hat * \hat j) + \rho \hat \omega
     \ \leftrightarrow \
     \left( \matrix -\hat * \hat j \\ \rho \hat \omega \endmatrix \right)
     \tag 6.6$$
and so the 3+1 decomposition of the Maxwell equations
$$d*F=-4\pi *j \tag 6.7$$
$$dF=0 \tag 6.8$$
and the continuity equation
$$d*j=0, \tag 6.9$$
respectively result in
$$\hat d \hat * \hat E + \hat y \wedge \hat * \hat B = 4\pi \rho \hat \omega
\tag 6.7a$$
$$\hat d \hat * \hat B - {\Cal L}_V \hat * \hat E - \hat a \wedge \hat *
  \hat B = 4\pi \hat * \hat j \tag 6.7b$$
$$\hat d \hat E + {\Cal L}_V \hat B - \hat a \wedge \hat E = 0 \tag 6.8a$$
$$\hat d \hat B - \hat y \wedge \hat E = 0 \tag 6.8b$$
and
$${\Cal L}_V (\rho \hat \omega ) + \hat d \hat * \hat j - \hat a \wedge
  \hat * \hat j =0\tag 6.10$$
In particular in the simplest situation, viz. for the irrotational ($\hat y=0$),
geodesic ($\hat a=0$) observer field $V$
(then $V={\partial}_t , \ \tilde V = dt$) we get
$$\hat d \hat * \hat E  = 4\pi \rho \hat \omega \tag 6.7a'$$
$$\hat d \hat * \hat B - {\Cal L}_{{\partial}_t} \hat * \hat E
                                         = 4\pi \hat * \hat j \tag 6.7b'$$
$$\hat d \hat E + {\Cal L}_{{\partial}_t} \hat B  = 0 \tag 6.8a'$$
$$\hat d \hat B  = 0 \tag 6.8b'$$
and
$${\Cal L}_{{\partial}_t} (\rho \hat \omega ) + \hat d \hat * \hat j
                                                             =0\tag 6.10'$$
(there is a simple rule to modify these equations to the case $\hat a\neq 0$
but still $\hat y=0$; cf. Appendix I).
\newline \indent
Since the equations \thetag {6.7a} - \thetag {6.10} are written in terms of
differential forms and standard well-behaved operations with respect to
integrals, one can readily write down their corresponding {\it integral
versions} : let spatial domains of necessary dimensions exist (2-dimensional
surface $\Cal S$, 3-dimensional volume $\Cal D$ - the latter case needs
$\hat y=0$, therefore we put $\hat y=0$ in the equations where the
integration over 3-dimensional domain is performed); then
$${\oint}_{\partial \Cal D} \hat * \hat E = 4\pi {\int}_{\Cal D}\rho
                                      \hat \omega \equiv 4\pi Q\tag 6.11a$$
$${\oint}_{\partial \Cal S} \hat * \hat B -
  \left.\frac{d}{d\tau} \right|_{0}
                            {\int}_{{\Phi}_{\tau} (\Cal S)} \hat * \hat E -
  {\int}_{\Cal S} \hat a \wedge \hat * \hat B =
  4\pi {\int}_{\Cal S} \hat * \hat j \tag 6.11b$$
$${\oint}_{\partial \Cal S} \hat E +
  \left.\frac{d}{d\tau} \right|_{0} {\int}_{{\Phi}_{\tau} (\Cal S)}
            \hat B - {\int}_{\Cal S} \hat a \wedge \hat E = 0 \tag 6.12a$$
$$ {\oint}_{\partial \Cal D}\hat B  = 0 \tag 6.12b$$
and
$$\left.\frac{d}{d\tau} \right|_{0}{\int}_{{\Phi}_{\tau} (\Cal D)}
                                                \rho \hat \omega  +
   {\oint}_{\partial \Cal D} \hat d \hat * \hat j -
   {\int}_{\Cal D} \hat a \wedge  \hat * \hat j =0\tag 6.13$$
where ${\Phi}_{\tau}$ is the (local) flow generated by $V$.
\newline \indent
The equations \thetag {6.7a} - \thetag {6.10} can be also expressed in more
familiar form, making use of 3-dimensional vector analysis operators div,
curl etc.; this is done in Appendix H (cf. \thetag {H.8} - \thetag {H.12}).
\newline \indent
Equivalently, if instead of $\thetag {6.7}$
$$\delta F = 4\pi j \tag 6.14$$
is used, $\thetag {6.7a}$, $\thetag {6.7b}$ are to be replaced by
$$\hat \delta \hat E - \hat * ( \hat y \wedge \hat * \hat B ) = 4\pi \rho
  \tag 6.14a$$
$$\hat \delta \hat B -\hat * {\Cal L}_V \hat * \hat E - \hat * ( \hat a
  \wedge \hat * \hat B ) = 4\pi \hat j \tag 6.14b$$
(they can be obtained directly by applying $\hat *$ on \thetag {6.7a},
\thetag {6.7b}, too.)
\newline \indent
The decomposition of the 4-{\it potential} 1-form
$$A\leftrightarrow
   \left( \matrix \phi \\ -\hat A \endmatrix \right) \tag 6.15$$
gives
$$\left( \matrix \hat E \\ -\hat B \endmatrix \right)
    \leftrightarrow F \equiv dA
    \leftrightarrow
    \left( \matrix -\hat d + \hat a & {\Cal L}_V \\ \hat y &
    \hat d \endmatrix \right)
    \left( \matrix \phi \\ -\hat A \endmatrix \right) =
    \left( \matrix -\hat d\phi + \phi \hat a -{\Cal L}_V\hat A
    \\ \phi \hat y -\hat d \hat A \endmatrix \right)  \tag 6.16$$
so that
$$\hat E = -\hat d\phi + \phi \hat a -{\Cal L}_V\hat A \tag 6.17$$
$$\hat B = \hat d \hat A -\phi \hat y \tag 6.18$$
Finally, the gauge transformation:
$$A\mapsto A' \equiv A+d\chi
  \leftrightarrow
  \left( \matrix \phi \\ -\hat A \endmatrix \right)  +
  \left( \matrix -\hat d + \hat a & {\Cal L}_V \\ \hat y &
                                 \hat d \endmatrix \right)
  \left( \matrix 0 \\ \chi \endmatrix \right) \tag 6.19$$
is
$$\phi \mapsto {\phi}' \equiv \phi + V\chi \tag 6.20$$
$$\hat A \mapsto {\hat A}' \equiv \hat A -\hat d \chi \tag 6.21$$
\vskip 1cm \noindent
{\bf VII. Conclusions and summary}
\vskip 0.5cm
\indent
In this article we presented a simple method of 3+1 decomposition of the
physical equations written in terms of differential forms on spacetime
$(M,g)$ with respect to a general observer field $V$.
\newline \indent
The method consists of the decomposition of both forms and operations on
them. The decomposition of forms is based technically on a simple identity
\thetag {2.4}, which can be interpreted in terms of projection operators on
${\Omega}^p(M)$. The decomposition of the operations on forms consists
first in the decomposition \thetag {4a.2} of the Hodge star operator
and then the decomposition of the exterior derivative $d$. Here the
formalism mimics the approach used standardly in the theory of connections
on principal bundle, viz. we first introduce the operator hor  (projecting
on the `` spatial part" of the form; its simple
realization is given by \thetag {3.2}) and then define the {\it spatial}
exterior derivative as $\hat d := \text{hor} \ d$ (the counterpart of the
{\it covariant} exterior derivative on principal bundle with connection).
The decomposition of $d$ is then given by \thetag {4b.5}. The essential
property of $\hat d$, which makes it a useful object, is the validity of
the {\it spatial Stokes formula} \thetag {4b.1}. It provides the usual link
between the differential and integral formulations of the physical laws
respectively. The language of differential forms on both 4 and 3+1 levels
turns out to be the most convenient tool for realization of this link,
since forms are the objects directly present under the integral signs.
\newline \indent
Let us mention, that also the quantities of physical interest which ``are not"
forms (energy-momentum tensor, Ricci and Einstein tensors,...) admit
description in terms of forms [4]; it is then possible to apply the
decomposition presented here also to them.
\vskip 1cm
\noindent
\leftline{\bf Acknowledgment}
\vskip 0.5cm
I would like to thank my wife \v Lubka for the patience.
\vskip 1cm
\noindent
{\bf Appendix A : Proof of \thetag{3.2}}
\vskip 0.5cm
The formula to be proved
$$(\text{hor} \ \alpha )(U, ..., W) \equiv \alpha (\text{hor} \ U, ...,
\text{hor} \ W) =
                          ( i_Vj_V \alpha )(U,...,W) \tag A.1$$
is $\Cal F$-linear $\Rightarrow$ it is enough to take either all vector fields
($U,...,W$) horizontal or one vertical (then V is enough) and the rest
horizontal. The former case means to check
$$\alpha (U, ..., W) = \alpha (U,...,W) - (\tilde V \wedge i_V\alpha)
                                              (U,...,W), \tag A.2$$
the latter case
$$0 = (i_Vi_Vj_V\alpha )(U,...,W) \tag A.3$$
Both are easily seen to hold.
\vskip 1cm
\noindent
{\bf Appendix B : Proof of \thetag{4a.2}}
\vskip 0.5cm
In general one has in any orthonormal right-handed frame by definition
$$* \undersetbrace p \to {e^a\wedge ...\wedge e^b} = \frac{1}{(n-p)!} \
{\eta}^{ac}...{\eta}^{bd} {\epsilon}_{c...de...f}e^e\wedge ... \wedge e^f
                                                              \tag B.1$$
Let
$$e^{0i...j} := e^0\wedge \undersetbrace {p-1} \to {e^i \wedge ...\wedge e^j}
  \hskip 1cm
  e^{k...l} := \undersetbrace p \to {e^k \wedge ...\wedge e^l} \tag B.2$$
be mixed and spatial basis p-forms in $(M^n;g;\undersetbrace n \to {+-...-})$
respectively and let us treat the orthogonal complement to $V\equiv e_0$ as
the Euclidean space with the signature (+...+) (its dimension is $n-1$).
Then
$$*e^{0i...j} = \frac{1}{(n-p)!}
  \undersetbrace {{(-1)}^{p-1}}
             \to {{\eta}^{00}{\eta}^{ii}...{\eta}^{jj}}
  \undersetbrace {((n-1)-(p-1))! \hat * e^{i...j}}
             \to {{\epsilon}_{0i...jk...l}e^{k...l}} =$$
$$ = {(-1)}^{p-1} \hat * e^{i...j} = \hat * \hat \eta e^{i...j} \tag B.3$$
and
$$* e^{k...l} = \frac1{(n-p)!}\ \undersetbrace (-1)^p \to{\eta^{kk}\dots
\eta^{ll}} \ \varepsilon_{k...la...b}\ e^{a...b} = $$
$$ = \frac{(-1)^p}{(n-p)!}\ (n-p)\ \varepsilon_{k...l0r...s}\ e^{0r...s} =$$
$$= \frac1{(n-p-1)!}\ \varepsilon_{0k...lr...s}\ e^{0r...s} = e^0\wedge \hat*\
  e^{k...l} \tag B.4$$
Then if
$$  \hat s \equiv \frac1{(p-1)!}\ \hat s_{i...j}\ e^{i...j} \hskip 2cm
                                                  (p-1)-\text{form}$$
$$ \hat r \equiv \frac1{p!}\ \hat r_{k...l}\ e^{k...l} \hskip 2cm
                                             p-\text{form} $$
and
$$ \alpha = e^0 \wedge \hat s + \hat r, \tag B.5$$
we have
 $$ * (e^0\wedge \hat s + \hat r) = \frac 1{(p-1)!}\ \hat s_{i...j}\ *
  e^{0i...j} + \frac1{p!}\ \hat r_{k...l}\ *e^{k...l} = e^0\wedge \hat *\hat r
  + \hat *\hat\eta\hat s \tag B.6$$
\vskip 1cm
\noindent
{\bf Appendix C : Interpretation of $\hat a$ and $\hat y$ in \thetag{4b.3}}
\vskip 0.5cm
Let
  $$ d\tilde V = \tilde V\wedge \hat a + \hat y $$
be the decomposition \thetag {4b.3} of the  2-form $d\tilde V$.Then according
to \thetag {2.9} one has
$$\hat a = i_Vd\tilde V
  = (\undersetbrace {\Cal L}_V\to{i_Vd+di_V})\tilde V
     - d\undersetbrace ||V||^2 = 1 \to{i_V\tilde V}
  = {\Cal L}_V\tilde V =$$
$$= {\Cal L}_V(g(V,.))
  = ({\Cal L}_Vg)(V,.) + g(\undersetbrace [V,V]=0\to{{\Cal L}_VV},.)
  = ({\Cal L}_Vg)(V,.) \tag C.1$$
  However, for the Levi-Civita connection one has in arbitrary coordinates
  $$ ({\Cal L}_Vg)_{ij} = V_{i;j} + V_{j;i} \tag C.2$$
$\Rightarrow$
  $$ (({\Cal L}_Vg)(V,.))_i = V^jV_{i;j} + V^jV_{j;i} = 
  (\nabla_V\tilde V)_i + (d\frac12 \undersetbrace 1 \to{||V||^2})_i$$
so that
  $$ \hat a = ({\Cal L}_Vg)(V,.) = \nabla_V\tilde V =
       \undersetbrace 0 \to{(\nabla_Vg)}(V,.) + g(\nabla_VV,.) \equiv g(a,.)
                                                                  \tag C.3$$
where
$$a := {\nabla}_VV\tag C.4$$
is the {\it acceleration field} corresponding to the observer field $V$; thus
  $$ \hat a = g(a,.) \equiv \tilde a \tag C.5$$
The 2-form $\hat y$:
\newline
According to the Frobenius theorem, $\hat y \neq 0$ means the nonintegrability
of the {\it horizontal} (3-space) {\it distribution}. This can be rephrased
as the
impossibility of the synchronization of the clocks within the 3-space for
the observers moving along $V$, too. Indeed, let
$$t:\Cal U \rightarrow \Bbb R $$
be a time function (coordinate) in a 4-region $\Cal U$, synchronized for
any two nearby space-related points, i.e.
\newline
1. $Wt = 0$ for any horizontal $W$ ($t$ is constant along the instantaneous
                                                                  3-space)
\newline
2. $Vt \equiv \chi > 0$ ( time increases along any observer's worldline).
\newline
The condition 1. can be rewritten also as
$$0=(\text{hor} \ W)t = \langle dt, \text{hor} \ W \rangle
                   = \langle \text{hor} \ dt, W \rangle
           \equiv    \langle \hat dt, W \rangle$$
for {\it any} $W$, i.e.
$$\hat d t = 0 \tag C.6$$
as a 1-form. According to \thetag {D.1}
$$0=\hat d t=dt - (Vt)\tilde V \equiv dt - \chi \tilde V$$
or
$$\tilde V = \psi dt \hskip 2cm \psi \equiv {\chi}^{-1} > 0 \tag C.7$$
Then
$$d\tilde V = d\psi \wedge dt = \frac{d\psi}{\psi}\wedge \psi dt = \tilde V
                                             \wedge (-\frac{d\psi}{\psi}) =$$
$$= \tilde V \wedge (-\frac{\hat d \psi + (V\psi)\tilde V}{\psi}) =
                                \tilde V \wedge (-\frac{\hat d \psi}{\psi})$$
Comparison with \thetag {4b.3} then gives
$$\hat y = 0 \hskip 1cm \hat a = -\frac{\hat d \psi}{\psi}\equiv -\hat d
  \Phi \hskip 1cm \psi \equiv e^{\Phi} \tag C.8$$
Thus $\hat y\neq 0$ is the obstacle for existence of a time function $t$
synchronized in 3-space and (if $\hat y = 0$)
$\Phi := \text{ln} \ \psi$ is the ``gravitational potential" ([6] , p.33).
\vskip 1cm
\noindent
{\bf Appendix D : The proofs of \thetag{4b.5}, and \thetag {4b.7}}.
\vskip 0.5cm
Let $\hat b$ be any spatial form, i.e. $i_V\hat b=0$; then
$$d\hat b  = (j_Vi_V + i_Vj_V)d\hat b
        = \tilde V \wedge \undersetbrace {\Cal L}_V - di_V\to{i_Vd}\hat b +
          \undersetbrace\hat d\to{i_Vj_Vd} \hat b =$$
$$= \tilde V\wedge{\Cal L}_V\hat b + \hat d\hat b - \tilde V\wedge d
                                    \undersetbrace 0\to{i_V\hat b} $$
so that on {\it spatial} forms
  $$ d\hat b = \tilde V\wedge {\Cal L}_V\hat b + \hat d\hat b \tag D.1$$
  Then on a {\it general} form $\alpha$
$$d\alpha = d(\tilde V \wedge \hat s + \hat r)
         = d\tilde V \wedge \hat s - \tilde V \wedge d\hat s + d\hat r =$$
$$ = (\tilde V \wedge \hat a + \hat y) \wedge \hat s - \tilde V
           \wedge (\tilde V \wedge \Cal L_V\hat s + \hat d \hat s) +
           \tilde V \wedge \Cal L_V \hat r + \hat d \hat r =$$
$$ = \tilde V \wedge (\Cal L_V \hat r - \hat d \hat s + \hat a \wedge
           \hat s) + (\hat d \hat r + \hat y \wedge \hat s) \tag D.2$$
where all forms in the brackets are already spatial. 
\newline
The computation of $\hat d \hat d$ : for arbitrary horizontal form $\hat b$
$$\hat d\hat d\hat b \ \overset(D.1)\to= \ \text {hor} \ d (d\hat b -
  \tilde V \wedge {\Cal L}_V\hat b) =  - \text {hor} \ (d\tilde V \wedge
  {\Cal L}_V\hat b - \tilde V\wedge d{\Cal L}_V  \hat b) \ \overset(3.7)\to=$$
$$= -\undersetbrace \hat y\to{\text {hor} \ d\tilde V}\wedge\undersetbrace
  {\Cal L}_V\hat b\to{\text {hor} {\Cal L}_V\hat b} + \undersetbrace 0
  \to{\text {hor} \ \tilde V}\wedge \text {hor} \ d{\Cal L}_V\hat b =
  -\hat y\wedge {\Cal L}_V \hat b \tag D.3$$
since
$$\text {hor} \ {\Cal L}_V\hat b\ =\ i_Vj_V(i_Vd + d\undersetbrace 0\to{i_V)
  \hat b} \ =\ i_V\undersetbrace 1-i_Vj_V\to{j_Vi_V}d\hat b\ =\ i_Vd\hat b -
  \undersetbrace 0 \to{i_Vi_V}j_Vd\hat b\ =\ (i_Vd + d\undersetbrace 0 
  \to{i_V)\hat b}\ =\ {\Cal L}_V \hat b \tag D.4$$
\vskip 1cm
\noindent
{\bf Appendix E : The volume expansion coefficient $\theta$}
\vskip 0.5cm
According to \thetag {4a.2}
$$\omega \equiv *1 = \tilde V \wedge \hat *1 \equiv \tilde V \wedge \hat
\omega$$
where $\hat \omega \equiv \hat *1$ is the {\it spatial} volume form.
The standard definition of the {\it volume expansion} coefficient $\theta$ is
([6], p. 9)
$$\theta := V^{\mu}{}_{;\mu} \equiv \nabla . V \equiv \text{div}V \tag E.1$$
Since
$${\Cal L}_V \omega = (\text{div}V) \omega \equiv \theta \omega $$
one can write
$$\theta \omega = \tilde V \wedge (\theta \hat \omega) = {\Cal L}_V\omega =
                                  ({\Cal L}_V\tilde V) \wedge \hat \omega +
                \tilde V \wedge {\Cal L}_V\hat \omega = (\text{cf. App.C})$$
$$= \hat a \wedge \hat \omega + \tilde V \wedge {\Cal L}_V \hat \omega$$
The first term vanishes (spatial 4-form), ${\Cal L}_V\hat \omega$ is
spatial (the end of App.D) so that
$${\Cal L}_V\hat \omega = \theta \hat \omega \tag E.2$$
which means that $\theta$ is the rate of change of {\it 3-volumes}
along the observer's wordline $\gamma$ (with $\dot \gamma = V$).
As an example ${\Cal L}_V(\rho \hat \omega) = (V\rho + \theta) \hat \omega$
(see \thetag {6.10}).
\vskip 1cm
\noindent
{\bf Appendix F : The identities resulting from $dd=0$}
\vskip 0.5cm
Applying $d$ on the decomposition \thetag {4b.3} and using \thetag {4b.5} one
obtains
$$0=dd\tilde V = d\tilde V \wedge \hat a - \tilde V \wedge d\hat a + d\hat
                                                                  y = $$
$$= \tilde V \wedge ({\Cal L}_V \hat y - \hat d \hat a) +
    (\hat d \hat y + \hat y \wedge \hat a)$$
so that $\hat a$ and $\hat y$ are always related by
$$\hat d \hat a = {\Cal L}_V \hat y \tag F.1$$
$$\hat d \hat y = - \hat y \wedge \hat a \tag F.2$$
In particular, for $\hat y=0$ we get
$$\hat d \hat a =0. \tag F.3$$
From \thetag {C.8} we even know, that
$$\hat a = -\hat d \Phi \tag F.4$$
in this case.
\newline \indent
Applying $dd=0$ on a spatial form $\hat r$ and taking into account \thetag
{F.1}, \thetag {F.2} we obtain another useful identity
$$[{\Cal L}_V,\hat d \ ] = - \hat a \wedge {\Cal L}_V \hskip 1cm \text{on} \
                                                 \hat \Omega (M) \tag F.5$$
(this shows that ``time" derivative ${\Cal L}_V$ and ``space" derivatives
hidden in $\hat d$ do not commute in general; they do commute, however, for
the geodesic observer field ($\hat a =0$)).
\vskip 1cm \noindent
{\bf Appendix G : Formal links with the theory of
                                          connections on principal bundles}
\vskip 0.5cm
The formulation used in this article resembles in many respects the theory of
connections on principal bundles [5]. There is a (right) action $R_g$ of a
Lie group G on the total space $P$, if we set $G \equiv (\Bbb R , +),\
P\equiv M$
(globally one needs $V$ to be {\it complete} for this) and the action is
identified with the flow
generated by $V$. The difference is, however, that the horizontal
distribution here is {\it not} $G$-invariant in general : since $\tilde V$
is the counterpart of the {\it connection form} $\omega$ - both define the
horizontal distribution via annihilation - and the group is 1-dimensional, the
$(\Bbb R,+)$ - invariance means ${\Cal L}_V\tilde V =0$. But
$${\Cal L}_V\tilde V \ \overset(C.3)\to= \ \hat a \neq 0 \hskip 1cm
                                       \text{in general} \tag G.1$$
(thus for {\it geodesic} ($\equiv non$accelerating) observer field there
{\it is} in fact $\Bbb R$-connection available).
\newline \indent
Many formulas here are very similar to those in the connection theory; e.g.
$$\hat d \hat b = d\hat b -\tilde V \wedge {\Cal L}_V \hat b \tag G.2$$
(cf. \thetag{D.1}) is the counterpart of the standard formula valid for
the computation of the {\it covariant} exterior derivative of the
horizontal form $\alpha$ of type $\rho$, viz.
$$D\alpha = d\alpha +{\rho}'(\omega)\dot \wedge \alpha \tag G.3$$
where $\omega$ is the connection form, $\rho$ a representation of $G$.
To see this more explicitly, one has to notice that the forms of type $\rho$
satisfy 
$${\Cal L}_{{\xi}_i} \alpha = - {\rho}'(E_i)\alpha \tag G.4$$
(${\xi}_i \equiv {\xi}_{E_i}$ being the fundamental field corresponding to the
basis element $E_i$ of the Lie algebra $\Cal G$ of $G$) and consequently
$${\rho}'(\omega)\dot \wedge \alpha
   = {\omega}^i\wedge {\rho}'(E_i)\alpha
   = -{\omega}^i\wedge {\Cal L}_{{\xi}_i} \alpha \tag G.5$$
or
$$D\alpha = d\alpha -{\omega}^i\wedge {\Cal L}_{{\xi}_i} \alpha \tag G.6$$
This formula is valid for {\it all horizontal} forms on the principal bundle
(not only of type $\rho$). For 1-dimensional group (as is the case here)
we have exactly the form \thetag {G.2}.
\newline \indent
In the same way one can see the similarity of
$$\hat d \hat d \hat b= -\hat y \wedge {\Cal L}_V \hat b\tag G.7$$
\thetag {4b.7} with the standard formula
$$DD\alpha = {\rho}'(\Omega)\dot \wedge \alpha
                       \equiv {\Omega}^i\wedge {\rho}'(E_i)\alpha \tag G.8$$
where $\Omega$, the curvature 2-form, is the counterpart of our $\hat y$ :
both encode the (non)integrability of the horizontal distribution and
consequently both are computed by the same rule, viz.
$$\hat y := \text{hor} \ d\tilde V \equiv \hat d \tilde V
  \hskip 1cm \text{versus} \hskip 1cm
  \Omega :=  \text{hor} \ d \omega \equiv D\omega.$$
\newline \indent
The counterpart of the {\it Bianchi identity} $DD\omega =0$ is
$$\hat d \hat d \tilde V = \hat d \hat y  \ \overset(F.2)\to= \
                                                 -\hat y\wedge \hat a$$
This is {\it not} zero in general, but it {\it is} zero for $\hat a=0$,
where $\tilde V$ {\it does} define a connection.
\vskip 1cm
\noindent
{\bf Appendix H : The Maxwell equations in the standard vector analysis
                                                             notations}
\vskip 0.5cm
We use the standard 3-dimensional Euclidean space relations (cf. definitions
in \thetag {6.2})
$$       \hat E = {\bold E}.d{\bold r} \hskip 0.5cm
         \hat B = {\bold B}.d{\bold S} \hskip 0.5cm
         \hat j = {\bold j}.d{\bold r} \hskip 0.5cm
         \hat y = {\bold y}.d{\bold S} \hskip 0.5cm
         \hat a = {\bold a}.d{\bold r} $$
$$\hat * \hat E = {\bold E}.d{\bold S} \hskip 0.5cm
  \hat * \hat B = {\bold B}.d{\bold r} \hskip 0.5cm
  \hat * \hat j = {\bold j}.d{\bold S} \hskip 0.5cm
  \hat * \hat y = {\bold y}.d{\bold r} \hskip 0.5cm
  \hat * \hat a = {\bold a}.d{\bold S} $$
$$\tag H.1$$
$$\hat y \wedge \hat * \hat B = (\bold y . \bold B )\hat \omega
                                \hskip 0.5cm
  \hat a \wedge \hat * \hat B = (\bold a \times \bold B ) . d\bold S
                                \hskip 0.5cm
  \hat a \wedge \hat * \hat j = (\bold a . \bold j )\hat \omega $$
$$ \hat y \wedge \hat E = (\bold y . \bold E )\hat \omega \hskip 0.5cm
   \hat a \wedge \hat E = (\bold a \times \bold E ).d\bold S$$
One can then introduce curl and div operations according to
$$       \hat d \hat E =: (\text{curl} \ \bold E ).d\bold S \tag H.2$$
$$\hat d \hat * \hat E =: (\text{div} \ \bold E )\hat \omega \tag H.3$$
and consequently then also
$$       \hat d \hat B =: (\text{div} \ \bold B )\hat \omega \tag H.4$$
$$\hat d \hat * \hat B =: (\text{curl} \ \bold B ).d\bold S \tag H.5$$
These definitions guarantee the validity of ``standard" integral formulas
like
$${\oint}_{\partial \Cal D} \bold E.d\bold S =
  {\int}_ {\Cal D} (\text{div} \ \bold E)\hat \omega \tag H.6$$
$${\oint}_{\partial \Cal S} \bold B.d\bold r =
  {\int}_ {\Cal S} (\text{curl} \ \bold B).d\bold S \tag H.7$$
as a consequence of the ``spatial" Stokes formula \thetag {4b.1}. Then
from \thetag {6.7a} - \thetag {6.10} we obtain
$$\text{div} \ \bold E + \bold y.\bold B = 4\pi \rho  \tag H.8$$
$$(\text{curl} \ \bold B ).d\bold S -{\Cal L}_V(\bold E.d\bold S) -
  (\bold a\times \bold B).d\bold S = 4\pi \bold j.d\bold S \tag H.9$$
$$(\text{curl} \ \bold E ).d\bold S +{\Cal L}_V(\bold B.d\bold S) -
  (\bold a\times \bold E).d\bold S = 0 \tag H.10$$
$$\text{div} \ \bold B - \bold y.\bold E =0 \tag H.11$$
and
$${\Cal L}_V (\rho \hat \omega ) + (\text{div} \ \bold j )\hat \omega -
  (\bold a.\bold j)\hat \omega =0 \tag H.12$$
or (cf. Appendix E)
$$V \rho +\theta \rho + \text{div} \ \bold j  - \bold a.\bold j =0 \tag H.13$$
as well as the corresponding integral versions
$${\oint}_{\partial \Cal D} {\bold E}.d{\bold S} =
  4\pi {\int}_{\Cal D}\rho \hat \omega \equiv 4\pi Q\tag H.14$$
$${\oint}_{\partial \Cal S} {\bold B}.d{\bold r} -
  \left.\frac{d}{d\tau} \right|_{0}
                            {\int}_{{\Phi}_{\tau} (\Cal S)} \bold E .d\bold S -
  {\int}_{\Cal S} (\bold a \times \bold B).d\bold S =
  4\pi {\int}_{\Cal S} \bold j . d\bold S \tag H.15$$
$${\oint}_{\partial \Cal S} \bold E.d\bold r +
  \left.\frac{d}{d\tau} \right|_{0} {\int}_{{\Phi}_{\tau} (\Cal S)}
                                                       \bold B.d\bold S -
  {\int}_{\Cal S} (\bold a \times \bold E) = 0 \tag H.16$$
$$ {\oint}_{\partial \Cal D} \bold B.d\bold S  = 0 \tag H.17$$
and
$$\left.\frac{d}{d\tau} \right|_{0}{\int}_{{\Phi}_{\tau} (\Cal D)}
                                                \rho \hat \omega  +
   {\oint}_{\partial \Cal D}  \bold j.d\bold S -
   {\int}_{\Cal D} (\bold a .\bold j) \hat \omega  =0\tag H.18$$
In the simplest situation, i.e. for irrotational ($\bold y=0$), geodesic
($\bold a=0$) observer field $V$ (then $V={\partial}_t , \tilde V = dt$) we get
$$\text{div} \ \bold E =  4\pi \rho \tag H.19$$
$$(\text{curl} \ \bold B ).d\bold S -{\Cal L}_{{\partial}_t}(\bold E.d\bold S)
                                    = 4\pi \bold j.d\bold S \tag H.20$$
$$(\text{curl} \ \bold E ).d\bold S +{\Cal L}_{{\partial}_t}(\bold B.d\bold S)
                                                = 0 \tag H.21$$
$$\text{div} \ \bold B =0 \tag H.22$$
and
$${\partial}_t \rho +\theta \rho + \text{div} \ \bold j  =0 \tag H.23$$
\vskip 1cm
\noindent
{\bf Appendix I : The equations $d\alpha = \beta$ and $\delta \alpha =
                                     \gamma$ for irrotational observer field}
\vskip 0.5cm
Let $V$ be an {\it irrotational} ($\hat y = 0$) observer field. Then (cf.
\thetag {C.8})
$$\hat a = -\hat d \Phi \hskip 1cm \Phi \equiv \text {ln} \psi \hskip 1cm
      \tilde V = \psi dt \hskip 1cm V = {\psi}^{-1} {\partial}_t \tag I.1$$
($e^{\Phi}\equiv \psi$ - {\it lapse function}, cf. [10]). Let us study the
equation of the structure
$$d\alpha = \beta . \tag I.2$$
If
$$\alpha \ \leftrightarrow \
           \left( \matrix \hat s  \\ \hat r \endmatrix \right)
           \hskip 1cm
  \beta  \ \leftrightarrow \
           \left( \matrix \hat S  \\ \hat R \endmatrix \right),
                                                             \tag I.3$$
we have
$$\left( \matrix -\hat d + \hat a & {\Cal L}_V \\ 0 &
    \hat d \endmatrix \right)
    \left( \matrix \hat s \\ \hat r \endmatrix \right) =
    \left( \matrix \hat S \\ \hat R \endmatrix \right)  \tag I.4$$
or
$$(-\hat d +\hat a)\hat s + {\Cal L}_V\hat r = \hat S \hskip 1cm
   \hat d \hat r = \hat R \tag I.5$$
Now
$$(-\hat d +\hat a)\hat s = -\hat d \hat s - \hat d \Phi \wedge \hat s
  = -e^{-\Phi}\hat d (e^{\Phi} \hat s)$$
$${\Cal L}_V \hat r = e^{-\Phi} {\Cal L}_{{\partial}_t} \hat r$$
so that we obtain
$$-\hat d (e^{\Phi} \hat s) + {\Cal L}_{{\partial}_t} \hat r =
   e^{\Phi} \hat S \hskip 1cm
   \hat d \hat r = \hat R \tag I.6$$
Thus we have the simple rule : the acceleration term $\hat a = -\hat d
\Phi$ manifests itself only through the replacement
$$\hat s \mapsto e^{\Phi}\hat s \equiv \psi \hat s \hskip 1cm
  \hat S \mapsto e^{\Phi}\hat S \equiv \psi \hat S \tag I.7$$
of the {\it upper} components of \thetag {I.3} (the lower ones being
unchanged) in the corresponding equations with $\hat a =0$, i.e. in
$$-\hat d \hat s + {\Cal L}_{{\partial}_t} \hat r =
   \hat S \hskip 1cm
   \hat d \hat r = \hat R \tag I.8$$
The similar analysis repeated for the equation
$$\delta \alpha
  = \gamma \hskip 1cm \text {where} \ \gamma \ \leftrightarrow \
    \left( \matrix \hat \frak S  \\ \hat \frak R \endmatrix \right) \tag I.9$$
shows that the replacement to be performed in the corresponding equations
with $\hat a =0$ is
$$\hat r \mapsto e^{\Phi}\hat r \equiv \psi \hat r \hskip 1cm
  \hat \frak R \mapsto e^{\Phi}\hat \frak R
                                   \equiv \psi \hat \frak R \tag I.10$$
i.e. only the {\it lower} components do change now.
\newline \indent
For the case of the Maxwell equations \thetag {6.14}, \thetag {6.8} and the
continuity equation \thetag {6.9} it results in the replacements
$$\hat E \mapsto e^{\Phi}\hat E \equiv \psi \hat E \hskip 1cm
  \rho \mapsto e^{\Phi}\rho \equiv \psi \rho  \tag I.11$$
in the {\it homogeneous} pair ($\rho$ is, however, trivial since it is not
present there),
$$  \hat B \mapsto e^{\Phi}\hat B \equiv \psi \hat B \hskip 1cm
  \hat j \mapsto e^{\Phi}\hat j \equiv \psi \hat j \tag I.12$$
in the {\it inhomogeneous} pair and
$$\hat j \mapsto e^{\Phi}\hat j \equiv \psi \hat j \tag I.13$$
in the continuity equation, i.e. the equations for $\hat a = -\hat d \Phi$
read (cf. [10], pp.18-19 and Appendix H here)
$$\hat d \hat * \hat E  = 4\pi \rho \hat \omega \hskip 2cm
  \hat d \hat * (e^{\Phi} \hat B) - {\Cal L}_{{\partial}_t} \hat * \hat E
  = 4\pi \hat * (e^{\Phi} \hat j) \tag I.14$$
$$\hat d (e^{\Phi} \hat E) + {\Cal L}_{{\partial}_t} \hat B  = 0 \hskip 2cm
  \hat d \hat B  = 0 \tag I.15$$
and
$${\Cal L}_{{\partial}_t} (\rho \hat \omega )
                          + \hat d \hat * (e^{\Phi} \hat j) = 0 \tag I.16$$
\vskip 1cm
\noindent
{\bf References}
\vskip 0.5cm
\noindent
$a)$ Present address : Department of Theoretical Physics, Co\-me\-nius
     University, Mlynsk\'a dolina F2, 842 15 Bratislava, Slovakia; E-mail:
     fecko\@fmph.uniba.sk
\newline \noindent
[1] K.S.Thorne, D.A.Macdonald  : Electrodynamics in curved spacetime :
     3+1 formulation, Mon. Not. R. astr. Soc. (1982) {\bf 198},339-343 +
     Microfiche
\newline \noindent
[2] Ch.W.Misner, J.A.Wheeler : Classical Physics as Geometry, Annals of
Physics, $\bold 2$, 525-603 (1957)
\newline \noindent
[3] J.Baez,J.P.Munian: Gauge Fields, Knots and Gravity, World Scientific,
                                                        1994
\newline \noindent
[4] I.M.Benn,R.W.Tucker: An Introduction to Spinors and Geometry with
            Applications in Physics, Adam Hilger, Bristol, 1989
\newline \noindent
[5] A. Trautman : Fiber Bundles, Gauge Fields, and Gravitation, in A.Held :
    General Relativity and Gravitation, Vol. 1, Plenum Press 1980
\newline \noindent
[6] G.F.R. Ellis : Relativistic cosmology, Carg\`ese Lectures in Physics,
                                                 Vol.6, 1-60, 1973
\newline \noindent
[7] N.Straumann : General Relativity and Relativistic Astrophysics,
               Springer - Verlag 1991, p.439
\newline \noindent
[8] Ch.W.Misner, K.S.Thorne, J.A.Wheeler : Gravitation, W.H.Freeman and
                                                         Company, Ex.22.6
\newline \noindent
[9] V.I.Arnold : Mathematical Methods of Classical Mechanics,
                                          Benjamin/Cummings Reading MA, 1978
\newline \noindent
[10] K.S.Thorne, R.H.Price, D.A.Macdonald :
                   Black Holes : The Membrane Paradigm , Yale Univ. Press 1986
\enddocument
\end